\documentclass[aps,pra,reprint,showpacs,amssymb,amsmath,longbibliography]{revtex4-1}
\usepackage{graphicx,amssymb,amsmath}

\def\be{\begin{equation}}
\def\ee{\end{equation}}
\def\ber{\begin{eqnarray}}
\def\eer{\end{eqnarray}}
\def\bern{\begin{eqnarray*}}
\def\eern{\end{eqnarray*}}

\def\0v{\mathbf{0}}
\def\1v{\mathbf{1}}
\def\2v{\mathbf{2}}
\def\3v{\mathbf{3}}

\begin{document}
\title{Cross-over between  collective and independent-particle excitations
in  quasi-2D electron gas with one filled miniband}

\author{Vladimir U. Nazarov}

\affiliation{Research Center for Applied Sciences, Academia Sinica, Taipei 11529, Taiwan}

\begin{abstract}
While it has been recently demonstrated that, for quasi-2D electron gas  (Q2DEG) with one filled miniband,
the dynamic exchange $f_x$ and Hartree $f_H$ kernels cancel each other in the low-density regime $r_s\rightarrow \infty$ (by half and completely, 
for the spin-neutral and fully spin-polarized cases, respectively), here we analytically show that the same happens at arbitrary densities  at short distances.
This motivates us to study the confinement dependence of the excitations in  Q2DEG. 
Our calculations unambiguously confirm that, at strong confinements, 
the time-dependent exact exchange  excitation energies approach the single-particle Kohn-Sham ones for the spin-polarized case,
while the same, but less pronounced, tendency is observed for spin-neutral Q2DEG.
\end{abstract}
\maketitle
\section{Introduction}
\label{intro}

As the density functional theory (DFT) \cite{Hohenberg-64,Kohn-65}  greatly facilitates
the study of the ground-state properties of many-body systems, substituting them with the single-particle ones, 
the time-dependent DFT (TDDFT) \cite{Runge-84,Gross-85} provides, in a similar way, a means to consistently study excitations.
Both theories are, in principle, exact, but, in practice, their  implementations rely on specific approximations to the exchange-correlation (xc) potentials.

In the case of TDDFT, the simplest and the widest used is the adiabatic local density approximation (ALDA),
which for the dynamic xc potential utilizes that of the homogeneous EG with local, in space and time, electron density.
ALDA xc potential suffers from many well-known deficiencies, including the violation of the asymptotic $-1/r$ law for finite systems and
not reproducing the $\alpha/q^2$ singularity of $f_{xc}$ in insulating and semiconducting crystals \cite{Botti-04}.
An attractive alternative to ALDA is given by the time-dependent exact exchange (TDEXX) theory \cite{Ullrich-95,Gorling-97}, 
which is free from the above shortcomings. Being, in most cases, an orbital-dependent theory, the TDEXX is, however, usually difficult
to implement in practical calculations 
(for notable exceptions, see, however, numerical implementations for linear response of crystalline silicon \cite{Kim-02}
and nonlinear dynamics of quasi-2D electron gas \cite{Wijewardane-08}).

New insights in the properties of the TDEXX have been recently achieved by finding an exact analytical solution, in both linear and nonlinear regimes,
for a specific model system, that of Q2DEG with one filled miniband \cite{Nazarov-17}.
In particular, important cancellations between the exchange $f_x$ and Hartree $f_H$ kernels of TDDFT 
(by half and completely, for the spin-neutral and fully spin-polarized cases, respectively) have been found in the low EG density limit.
This leads to a conclusion that, in that limit, the spin-polarized quasi-2DEG responds as a collection of independent particles, 
rather than as a system supporting collective (interband plasmonic) excitations, the latter conventionally assumed \cite{Ullrich}.

The present work is devoted to the demonstration and discussion of a new observation 
that the same cancellations as well as their consequences for the quantum dynamics
occur in Q2DEG with
one filled miniband at {\it arbitrary density} but in the case of {\it strong confinement}.

This paper  is organized as follows:
In Sec.~\ref{Q2DEG} we recall the analytical TDEXX solutions for the exchange potential and kernel of  Q2DEG with one filled miniband.
In Sec.~\ref{Dep} we present results of calculations of the  excitations of the EG, demonstrate their confinement dependence, 
and discuss the  physical meaning of the new findings.
Conclusions are presented in Sec.~\ref{Concl}.

\section{TDEXX of quasi-2DEG with one filled miniband}
\label{Q2DEG}

\begin{figure}[h!]
\includegraphics[width= 1 \columnwidth, trim= 0 30 150 20, clip=true]{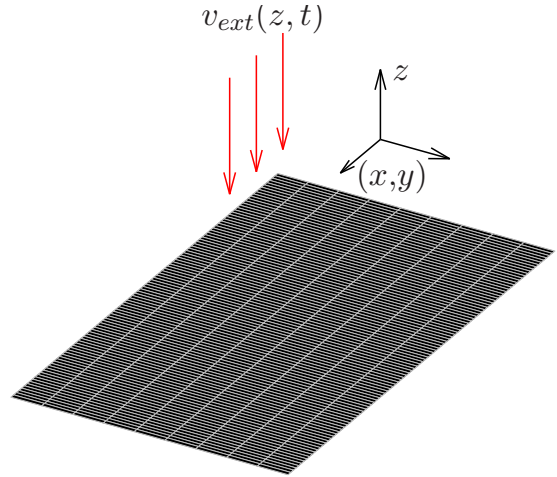}
\caption{\label{syse}
Schematics of the Q2DEG under the action of a time-dependent external potential.}
\end{figure}

As schematized in Fig.~\ref{syse}, we consider Q2DEG, 
which is homogeneous in the $xy$ plane and confined in the $z$ direction by a static external potential $v_{ext}(z)$.
A time-dependent external field along the $z$-axis, defined by the potential $v_{ext}(z,t)$, is applied to this system causing excitations of the  interacting EG.

It has been proven it Ref.~\cite{Nazarov-17} that the {\it fully dynamic} TDEXX potential in this setup experienced by an electron 
of the $z$-spin projection $\sigma$ is given by
\begin{equation}
\begin{split}
  v^\sigma_x(z,t)   = -\frac{1}{n^\sigma_{2D}}
    \int    \frac{F_2(k_F^{\sigma}|z -  z'|)}{|z  - z'|}  n^\sigma(z',t) d z' ,
\end{split}
\label{main152}
\end{equation} 
where  $n^\sigma(z,t)$ is the spin-density,
\begin{equation}
F_2(u)=1+\frac{ L_1(2 u)-I_1(2 u)}{u},
\label{F2D}
\end{equation}
$L_1$ and $I_1$ are the first-order modified Struve and Bessel functions \cite{Prudnikov,Mathematica}, respectively, 
$n^\sigma_{2D}=\int_{-\infty}^\infty n^\sigma(z,t) d z$ is the 2D spin-density, which does not change during the time-evolution,
and $k_F^{\sigma}=\sqrt{4\pi n^\sigma_{2D}}$ is the corresponding 2D Fermi radius.
Accordingly, the dynamic exchange kernel of this system is
\begin{equation}
f^{\sigma\sigma'}_x(z,z',\omega)=
-\frac{1}{n^\sigma_{2D}} \frac{F_2(k_F^{\sigma}|z -  z'|)}{|z  - z'|}\delta_{\sigma \sigma'}.
\label{fx} 
\end{equation}
As pointed out in Ref.~\cite{Nazarov-17}, although $f_x$ of Eq.~(\ref{fx}) is frequency-independent, Eqs.~(\ref{main152}) and (\ref{fx})
are exact results in TDEXX for this system and the excitation geometry, with no adiabatic approximation invoked.

In this work we will focus attention at the short-range behavior of $f_x$, i.e., at the regime of $|z -  z'| \ll (k_F^{\sigma})^{-1}$.
Obviously, the expansion of Eq.~(\ref{fx}) in the series in this case is the same as  in Ref.~\cite{Nazarov-17} for small $k_F^{\sigma}$, 
and it results in
\begin{equation}
f^{\sigma\sigma'}_x(z,z',\omega)\approx
\left( - \frac{32}{3 k_F^\sigma} + 2 \pi |z-z'| \right) \delta_{\sigma\sigma'}.
\label{fxexp} 
\end{equation}
Noting again that the first term in Eq.~(\ref{fxexp}) is a constant, which does not play a role,  comparing to the Hartree kernel,
which for Q2DEG is
\begin{equation}
f^{\sigma\sigma'}_H(z,z')  = - 2 \pi |z-z'|, 
\label{fH}
\end{equation}
and taking note of the expression for the interacting-particle density response function \cite{Gross-85}
\begin{equation}
\begin{split}
\left(\chi^{-1}\right)^{\sigma\sigma'} (z,z',\omega)  &=  
\left(\chi^{-1}_s\right)^{\sigma\sigma'}   (z,z',\omega) - f_H(z,z') \\
&- f^{\sigma\sigma'}_x(z,z',\omega),
\end{split}
\label{chchs}
\end{equation}
where $\chi_s$ is the KS density-response function, we conclude that in the short-range regime, as previously in the low-density one, exchange kernel cancels the Hartree one completely for fully spin-polarized Q2DEG and by half in the spin-neutral case.

Having realized this important property, we can expect that it will clearly manifest itself in EG strongly confined by an external potential.
We will see in the next section that it really is the case.

\section{Dependence of the many-body excitations of Q2DEG on the EG's confinement}
\label{Dep}

To follow the framework of Ref.~\cite{Nazarov-17} as close as possible, 
we keep the confining potential being that of the strictly
2D uniform positively charged sheet at $z=0$. 
To vary the confinement, we, however, set the 2D positive charge density $n_s^+$ greater than the electron density
$n_s$, our system being positively charged. Accordingly, we consider $r_s^+ < r_s$.\footnote{At $r_s^+> r_s$ the system is unstable, at least within EXX.}

\begin{figure}[h!]
\includegraphics[width= 1 \columnwidth, trim= 85 0 10 0, clip=true]{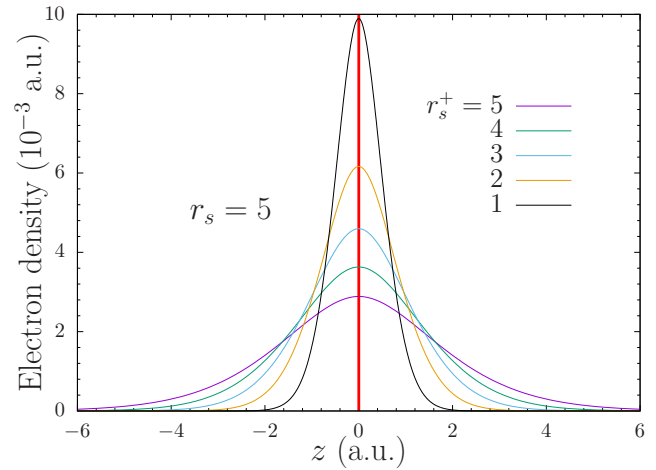}
\caption{\label{pol_n}
Ground-state EXX electron density distribution of the fully spin-polarized quasi-2DEG with the density parameter $r_s=5$.
The confining potential is that of the strictly 2D positive charge in the  $z=0$ plane with the density  corresponding to the parameter $r_s^+$. 
}
\end{figure}

In Fig.~\ref{pol_n} the ground-state electron density  calculated self-consistently using the exact analytical static EXX potential \cite{Nazarov-16-2}
is plotted for $r_s=5$ and for several values of $r_s^+$ in the case of the fully spin-polarized Q2DEG.
The EG confinement ranges can be clearly judged from this figure. 
\begin{figure}[h!]
\includegraphics[width= 1 \columnwidth]{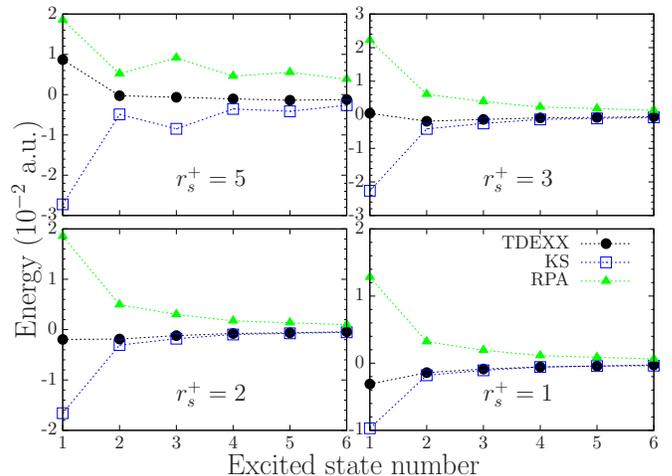}
\caption{\label{pol}
TDEXX, KS, and RPA excitation energies of the fully spin-polarized quasi-2DEG of the density parameter $r_s=5$, confined
with the strictly 2D uniform positive charge of the density parameter $r_s^+$.
For better visualization, each point is presented relative to the arithmetic mean of the TDEXX, KS, and RPA values.
}
\end{figure}

Our next step is to calculate the many-body excitation energies for the systems with different $r_s^+$, which are plotted in Fig.~\ref{pol}
for the TDEXX, random phase approximation (RPA) [setting $f_x=0$ in Eq.~(\ref{chchs})], and KS [setting $f_x=f_H=0$ in Eq.~(\ref{chchs})] theories.
We do this using Eq.~(\ref{chchs}) with the kernels of Eqs.~(\ref{fx}) and (\ref{fH}) and with $\chi_s$ calculated analytically by the method of Ref.~\cite{Nazarov-17},
but with $r_s^+\ne r_s$. A tendency of the many-body excitation energy getting closer to the single-particle KS values with the decreasing $r_s^+$, i.e., 
with the increasing confinement of the EG, can  be clearly seen from this figure.
We note that this is not the case for the charge-neutral system (upper left panel in Fig.~\ref{pol}) due to the insufficient confinement at $r_s^+=r_s$ (the broadest curve
in Fig.~\ref{pol_n}).
\begin{figure}[h!]
\includegraphics[width= 1 \columnwidth, trim= 0 0 0 0, clip=true]{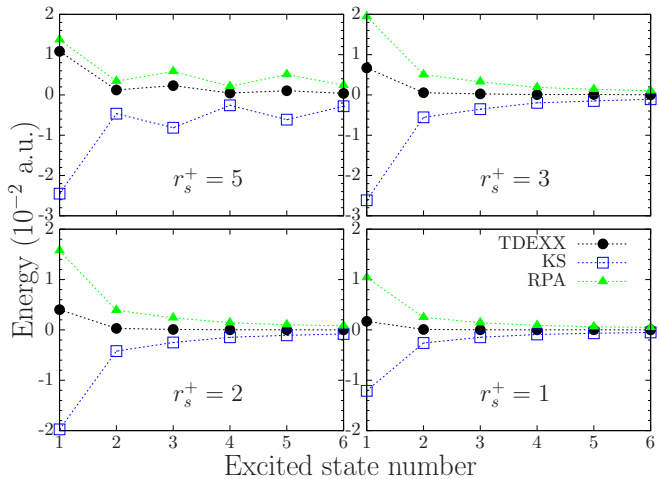}
\caption{\label{unpol}
The same as Fig.~\ref{pol}, but for the spin-neutral Q2DEG.
}
\end{figure}
Figure \ref{unpol} shows results of the same calculations, but for the spin-neutral case. Here the many-body excitation energies do not
tend to the KS values with the confinement increase, but are rather found between the latter and the RPA results.
This is in accordance with $f_x$ canceling only a half of $f_H$  in the spin-neutral case.\footnote{We do not plot the ground-state electron densities of the sin-neutral EG,
since they are very close to the spin-polarized ones.}

\section{Conclusions}
\label{Concl}
Within linear-response TDDFT, at the level of the time-dependent exact exchange, 
we have considered the role of the system confinement in many-body excitations of quasi-2D electron gas with one occupied miniband. 
We have taken advantage of the analytical solutions for TD exact exchange potential $v_x$ and exchange kernel $f_x$ 
recently obtained for excitations with a longitudinal TD  perturbations perpendicular to the EG layer.

Having analytically established that the exchange kernel $f_x$ cancels the Hartree one $f_H$ at short distances, completely and by half,
for fully spin-polarized and spin-neutral EG, respectively, we had anticipated this property's direct relevance to the nature of 
the Q2DEG response depending on its confinement. 
Upon conducting accurate calculations of the many-body excitation energies, we have found that,
in accordance to our expectation, at stronger confinements  the many-body excitation energies converge to the Kohn-Sham ones
in the fully spin-polarized case. For weaker confinements, these energies lie between the KS and RPA values.

These observations unambiguously suggest that a strongly confined spin-polarized electron gas with one filled miniband
responds to the perpendicular  perturbation as a gas of independent particles. On the other hand, at weaker confinements this system
exhibits collective behavior.

For the spin-neutral case, due to only a partial cancellation between $f_x$ and $f_H$, the above behavior is less pronounced,
the many-body excitation energies still differing from the KS ones at strong confinements. 
We, however, argue that even in the spin-neutral case the plasmonic nature 
of the excitations \cite{Ullrich} at strong confinements can be taken with reservations, since it is only a {\it half of the Hartree potential} 
which drives the system, while  the Hartree potential is the main origin of a conventional plasmons (cf. the classical self-sustained
charge-density oscillations \cite{Pines-63}).

Considering possible future generalizations of our results, we note that the assumption of one filled miniband is not restrictive in the context of our study at all,
since at a sufficiently strong perpendicular confinement only the first miniband will be occupied in any Q2DEG.
The inclusion of correlations and going beyond uniformity in the layer plane present real challenges, to be addressed in the future.

\acknowledgments
Discussion with C. A. Ullrich is highly appreciated.
Support from the Ministry of Science and Technology, Taiwan, Grant  No. 106--2112--M--001--021, is acknowledged.

%

\end{document}